\documentclass[10pt,twocolumn,letterpaper]{article}

\usepackage{iccv}
\usepackage{times}
\usepackage{epsfig}
\usepackage{graphicx}
\usepackage{amsmath}
\usepackage{amssymb}
\usepackage{color}
\usepackage{booktabs}
\usepackage{floatrow} 
\usepackage{subcaption}
\usepackage{makecell}

\usepackage[breaklinks=true,bookmarks=false]{hyperref}

\iccvfinalcopy 

\def\iccvPaperID{****} 

\ificcvfinal\fi

\begin{document}

\title{Robust Core-Periphery Constrained Transformer for Domain Adaptation}

\author{Xiaowei Yu\\
MS\&T\\
\and
Zeyu Zhang\\
UT-Arlington\\
\and
Dajiang Zhu\\
UT-Arlington\\
\and
Tianming Liu\\
University of Georgia\\
}

\maketitle
\ificcvfinal\thispagestyle{empty}\fi

\begin{abstract}
Unsupervised domain adaptation (UDA) aims to learn transferable representation across domains. Recently a few UDA works have successfully applied Transformer-based methods and achieved state-of-the-art (SOTA) results. However, it remains challenging when there exists a large domain gap between the source and target domain. Inspired by humans' exceptional transferability abilities to adapt knowledge from familiar to uncharted domains, we try to apply the universally existing organizational structure in the human functional brain networks, i.e., the core-periphery principle to design the Transformer and improve its UDA performance. In this paper, we propose a novel brain-inspired robust core-periphery constrained transformer (RCCT) for unsupervised domain adaptation, which brings a large margin of performance improvement on various datasets. Specifically, in RCCT, the self-attention operation across image patches is rescheduled by an adaptively learned weighted graph with the Core-Periphery structure (CP graph), where the information communication and exchange between image patches are manipulated and controlled by the connection strength, i.e., edge weight of the learned weighted CP graph. Besides, since the data in domain adaptation tasks can be noisy, to improve the model robustness, we intentionally add perturbations to the patches in the latent space to ensure generating robust learned weighted core-periphery graphs. Extensive evaluations are conducted on several widely tested UDA benchmarks. Our proposed RCCT consistently performs best compared to existing works, including 88.3\% on Office-Home, 95.3\% on Office-31, 90.7\% on VisDA-2017, and 46.0\% on DomainNet.
\end{abstract}

\section{Introduction}

\begin{figure*}[htp]
\begin{center}
\includegraphics[scale=0.34]{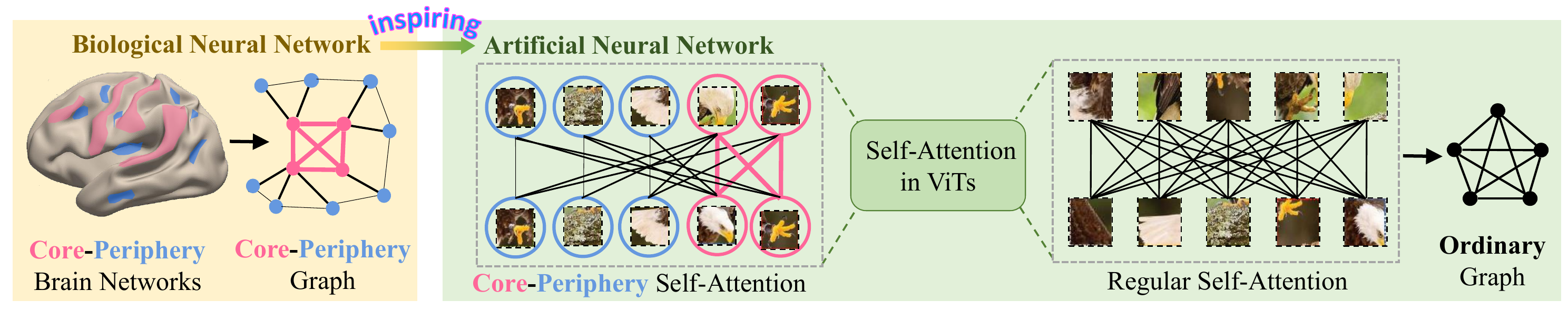}
\end{center}
\caption{The Core-Periphery principle in brain networks inspires the design of ANNs. The Core-Periphery structure broadly exists in brain networks, with a dense “core” of nodes (pink) densely interconnected with each other and a sparse “periphery” of nodes (blue) sparsely connected to the core and among each other. Inspired by this principle of BNN, we aim to instill the Core-Periphery structure into the self-attention mechanism and propose a new RCCT model.} 
\label{BG}
\end{figure*}

Deep neural networks (DNNs) made breakthroughs in various application fields due to their powerful automatic feature extraction capabilities. However, such impressive success usually needs great amounts of labeled data which can not be realized in the real case because of considerable time and expensive labor forces. Fortunately, unsupervised domain adaptation (UDA) \cite{Wilson20} techniques can leverage rich labeled data from the source domain and transfer knowledge from the source domain to the target domains with no or limited labeled examples. The key point of UDA is to find the discriminant and domain-invariant features from the labeled source domain and the unlabeled target domain in the common latent space. Along with more and more resources devoted to domain adaption research, the past decades have witnessed many UDA methods proposed and evolved \cite{Ganin15} \cite{Liang20} \cite{Long18} \cite{Rui18}\cite{Zhang19}. 

Recently, the self-attention mechanism and its variant vision transformer (ViT)~\cite{dosovitskiy2020image} have received growing interest in the vision community. Distinguished from Convolutional Neural Networks (CNN) that acquire information on local receptive fields of the given image, ViT models take advantage of the self-attention mechanism and, therefore, can obtain long-range dependencies among patch features through a global view. Specifically, for the vision transformer and its variants, each image is divided into a series of non-overlapping fixed-size patches, which are further projected into the latent space as patch embeddings/tokens and concatenated with position embeddings. A class token is prepended to the patch tokens, serving as the representation of the entire image. These patch tokens and the class token are delivered into a specific number of transformer layers to learn visual representations of the input image. Due to the superiority of the self-attention mechanism in global content capture, ViT and its variants have obtained impressive performance on kinds of vision tasks, such as image classification\cite{dosovitskiy2020image}, video understanding\cite{Girdhar19video} \cite{Neimark21video}, object detection\cite{carion2020end} \cite{Wang21}, and content segmentation\cite{zheng2021rethinking} \cite{liu2021swin}.

However, the existing methods are all artificial neural network (ANN) driven structures, including variants of CNNs in conjunction with advanced techniques, such as adversarial learning, or the newest structures of Transformers combined with effective techniques like self-refinement \cite{Sun22} \cite{Yang23} \cite{Xu22} \cite{XuICCV19}. More and more studies have found that the best-performing ANNs surprisingly resemble biological neural networks (BNN), which indicates that ANNs and BNNs may share common principles to achieve optimal performance in either machine learning or cognitive tasks \cite{You20}~\cite{YuCPVIT2023}. Inspired by the studies in information communication in brain networks, in this work, we aim to proactively instill the organizational principle of Core-Periphery structure in BNNs to improve the domain adaptation ability of ANNs. The concept of the Core-Periphery functional brain network is illustrated in Figure \ref{BG}, where the connections between cores are much denser and stronger than the counterparts between peripheries.

Aiming to bring brain-inspired priors into the ANNs, in this work, we propose a novel robust core-periphery constrained transformer (RCCT) for unsupervised domain adaptation. RCCT takes a vision transformer as the backbone network and manipulates the strength of self-attention under the core-periphery constraints so that the information communication and exchange among the core patches are more effective and efficient while weakening the unimportant information flows among the periphery patches. Moreover, RCCT has two key components that lead to its excellent performance, one is the core-periphery principle guided self-attention, and the other is the robust adaptive core-periphery graph generation.

We conclude our contributions as follows:

\noindent• We develop a novel UDA solution RCCT, which proactively installs the brain-inspired core-periphery principle to manipulate the connection strength of self-attention in the vision transformer to boost its strong transferable feature representation.

\noindent• We utilize the predictions on dual-domain perturbed data for generating the robust core-periphery graph. It makes the model adaptively learn the domain-invariant core patches and domain-specific periphery patches.

\noindent• RCCT is one of the pioneers in exploring the vision transformer for unsupervised domain adaptation. Also, our proposed RCCT demonstrates that the brain-inspired vision transformer has its superiority by showing SOTA results on the widely tested datasets, including 88.3\% on Office-Home, 95.0\% on Office-31, 90.7\% on VisDA-2017 and 46.0\% on DomainNet.

\section{Related Works}
\subsection{Core-Periphery Structure}
The Core-Periphery structure is a fundamental network signature that is composed of two qualitatively distinct components: a dense “core” of nodes strongly interconnected with one another, allowing for integrative information processing to facilitate the rapid transmission of messages, and a sparse “periphery” of nodes sparsely connected to the core and among each other \cite{gallagher2021clarified}. The Core-Periphery pattern has helped explain a broad range of phenomena in network-related domains, including online amplification\cite{barbera2015critical}, cognitive learning processes \cite{bassett2013task}, technological infrastructure organization \cite{alvarez2005k, carmi2007model}, and critical disease-spreading conduits \cite{kitsak2010identification}. All these phenomena suggest that the Core-Periphery pattern may play a critical role to ensure the effectiveness and efficiency of information exchange within the network. 

\subsection{Unsupervised Domain Adaptation}
UDA aims to learn transferable knowledge across the source and target domains with different distributions \cite{Pan09Survey} \cite{Ying18Transfer}. There are mainly two kinds of DNNs in terms of deep neural networks, which are CNN-based and Transformer-based methods \cite{Sun22} \cite{Yang23} \cite{Xu22} \cite{Liang20}. Various techniques for UDA are adopted on these backbone architectures. For example, the discrepancy techniques measure the distribution divergence between source and target domains \cite{Long18} \cite{Sun18Coral} \cite{Tzeng14Deep}. Adversarial adaptation discriminates domain-invariant and domain-specific representations by playing an adversarial game between the feature extractor and a domain discriminator \cite{Ganin15} \cite{Yang23}.

\subsection{Vision Transformer}
Vision Transformer (ViT) \cite{dosovitskiy2020image} is the first work that applies the transformer structure for image classification. However, there have been few applications of vision transformers for domain adaptation. Notably, some explorations \cite{Sun22} \cite{Yang23} \cite{Xu22} have been recently reported. Specifically, CDTrans \cite{Xu22} applies cross-attention to source-target image pairs. TVT \cite{Yang23} discriminates patch tokens and the class token with a local and a global discriminator. SSRT\cite{Sun22} utilizes bi-directional self-refinement on perturbed target domain and uses a safe training strategy. Our method is different from the existing works, where we add brian-inspired core-periphery constraints on the self-attention to manipulate the information communications among the patch tokens. Besides, we purposely add perturbations on the core-periphery graph generation to obtain a robust CP graph.

\begin{figure*}[ht]
\begin{center}
\includegraphics[scale=0.3]{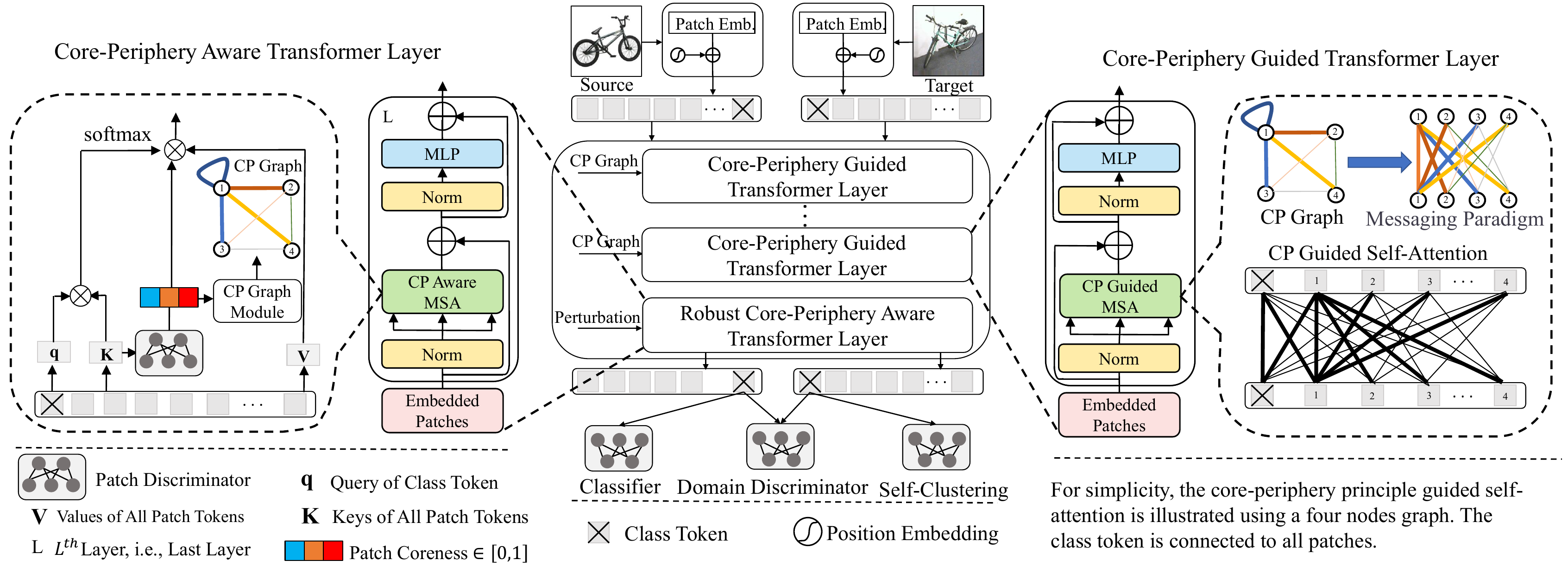}
\end{center}
\caption{ The overview of the proposed RCCT framework. In RCCT, source and target images are divided into non-overlapping fixed-size patches which are linearly projected into the latent space and concatenated with positional information. A class token is prepended to the image patches. The image patches and class token are delivered into a transformer encoder whose last layer is leveraged by a robust core-periphery aware layer, and the self-attention in previous layers is rescheduled by the adaptively core-periphery graph learned in the last layer. The robust core-periphery graphs are learned with deliberate perturbations. Domain-invariant/Domain-specific feature learning is therefore contained in core patches/periphery patches. Adversarial domain adaptation is accomplished by patch-level and global-level discriminators. The head classifier and self-clustering module are responsible for source domain images and target domain images classification, separately.}
\label{model}
\end{figure*}

\section{Methods}
\subsection{Preliminaries}
In UDA, the images set in the labeled source domain are represented as $D_{s}\left \{ \left ( x_{i}^{s},y_{i}^{s}   \right )  \right \}_{i=1}^{n_{s} } $, where $x_{i}^{s}$ are the images, $y_{i}^{s} $ are the corresponding labels, and $n_{s}$ are the number of samples. The target domain is represented as $D_{t}\left \{ \left ( x_{i}^{t}  \right )  \right \}_{j=1}^{n_{t} } $ with $n_{t}$ samples and no labels. UDA solutions aim to learn domain-invariant (core features) and domain-specific (periphery features) features to minimize the domain discrepancy, so as to obtain desired prediction performance on the unlabeled target data. 

The common practice is to design an objective function that jointly learns feature embeddings and a classifier. The objective function is formulated as 
\begin{equation}\label{GeneralObject}
min\{ L_{CE} \left \{ x_{s}, y_{s}   \right \} +\alpha L_{dis} \left \{ x_{s}, x_{t}   \right \} \}
\end{equation}
where $L_{CE}$ is the standard cross-entropy loss supervised in the source domain, $L_{dis}$ is a transferred loss with implementations of various solutions, and $\alpha$ is used to control the importance of $L_{dis}$. 

\subsection{ Methodology}

We aim to manipulate the strength of self-attention among dominant-invariant and dominant-specific patches by adaptively learning robust core-periphery graphs with perturbed data on the source and target domains. Figure \ref{model} illustrates the whole framework of our proposed RCCT. Since the source and target images are trained together, the source and target domain data are shown here. The network consists of a vision transformer backbone, a domain discriminator for the class token, a patch discriminator for patch tokens, a head classifier, a self-clustering, and a core-periphery graph generation model. For images of each domain, the Patch Embedding layer linearly mapped them into a token sequence including a special class token and image tokens.

For the initial epoch, since we do not know the coreness of each image patch. Therefore, we initiate with a unweight complete graph to guide self-attention. With each iteration, the patch discriminator in the robust core-periphery aware transformer layer will asses the coreness of each image patch. The patch discriminator also encourages the class token in the last transformer layer to focus on dominant-invariant features in core patches and contempt the dominant-specific features in periphery nodes. Then the CP graph module will generate a CP graph according to the patch coreness. In the meantime, we add perturbation to the core-periphery aware transformer layer to ensure its robustness to potential fluctuations. From the second epoch, the transformer layers before the robust CP aware layer will adopt the generated CP graph to reschedule the self-attention to strengthen the information communication among core patches (dominant-invariant) and weaken the connection among periphery patches (dominant-specific). 

The classifier takes the class token of the source domain images and outputs label prediction. The domain discriminator takes the output class tokens of the source and target domain to be aligned in the latent space by playing a two-player min-max game with the feature extractor. The self-clustering module enforces the aligned features of different classes of target-domain images to be clustered and separable.

\subsection{Robust Core-Periphery Aware Transformer Layer}
As shown in Figure \ref{model}, we introduce the Robust Core-Periphery Aware Transformer Layer that takes advantage of the intrinsic merits of ViT,
i.e., self-attention mechanisms and sequential patch tokens.
\subsubsection{Core-Periphery Aware Module}
The patch tokens correspond to partial regions of the image and capture visual features as fine-grained local representations. Existing work \cite{Yang23} shows that the patch tokens are of different semantic importance, in this work, we define the coreness of the core-periphery principle to index the importance of patches, higher coreness patches are more likely to correspond to the domain-invariant patches, whereas lower coreness patches corresponding to domain-specific patches. Core-periphery aware layer aims at learning different coreness indices to those patch tokens for two purposes. First, to encourage the global image representation, i.e., the class token in the last layer, to attend to core tokens. Second, to strengthen the information communications among the core tokens and weaken the information among periphery tokens by rescheduling the self-attention under the guidance of the core-periphery graphs generated via the coreness of each patch token.

To obtain the coreness of patch tokens, we adopt a patch-level domain discriminator $D_{l} $ to evaluate the local features by optimizing:
\begin{equation}\label{PatchLevelloss}
L_{pat}\left ( x^{s},x^{t} \right )  = -\frac{1}{nP} \sum_{x_{i} \in D} \sum_{p=1}^{P} L_{CE} \left ( D_{l} \left ( G_{f} \left ( x_{ip}^{*}  \right )   \right ) , y_{ip}^{d}  \right ) 
\end{equation}
where $P$ is the number of patches, $D=D_{s}\cup D_{t}$, $G_{f}$ is the encoder for feature learning, implemented as ViT, $n = n_{s} + n_{t}$, is the total number of images of the source and target domain, the superscript $*$ denotes a patch from either source or target domain, $x_{ip}^{*}$ represents the $p$th of the $i$th image, $y_{ip}^{d}$ denotes the domain label of the $p$th token of the $i$th image, i.e., $y_{ip}^{d} = 1$ means source domain, else the target domain. $D\left ( f_{ip}  \right )  $ gives the probability of the patch belonging to the source domain. During the training process, $D_{l}$ tries to discriminate the patches correctly, assigning 1 to patches from the source domain and 0 to those from the target domain, while $G_{f}$ combats such circumstances.

Empirically, patches that can easily deceive the patch dominator (e.g., $D_{l}$ is around 0.5) is more likely to be domain-invariant across domains and should be given a higher coreness. Therefore, we use $C\left (  f_{ip} \right ) = H\left ( D_{l}\left ( f_{ip}  \right )   \right ) \in \left [ 0, 1 \right ] $ to measure the coreness of $r$th token of $i$th image, where $H\left ( \cdot  \right ) $ is the standard entropy function. The explanation for the coreness is that by assigning an index to different patches, the model separates an image into domain-invariant representations and domain-specific representations, and the information communication from domain-specific features is softly suppressed. The generated core-periphery graph is then formulated as:
\begin{equation}\label{CPgraphGeneration}
M_{cp}= \frac{1}{BH}  \sum_{h=1}^{H}\sum_{b=1}^{B}      \left [   \left [ C\left (  f_{ip} \right )  \right ]^{T}  C\left (  f_{ip} \right ) \right ] _{\times }
\end{equation}

\begin{equation}\label{CPGraphRegularization}
M_{cp}\left ( i ,j\right )=  \begin{cases}
sqrt(M\left ( i ,j\right ))  & \text{ if } M\left ( i ,j\right )\ge 0.5 \\
square(M\left ( i ,j\right ))  & \text{ if } M\left ( i ,j\right )< 0.5
\end{cases}
\end{equation}
where $T$ means transpose of the matrix, $B$ is the batch size, $H$ is the number of heads, $\left [ \cdot  \right ] _{\times } $ means no gradients back-propagation for the adjacency matrix of the generated core-periphery graph, $sqrt(\cdot )$ and $square(\cdot )$ are the square root and square operations, respectively. The $sqrt(\cdot )$ and $square(\cdot )$ operations make the core-periphery property more apparent in the CP graph. The mask matrix $M_{cp}$ is the adjacency matrix of the core-periphery graph, and it defines the connection strength of the patch pairs. The connection strength among those patches with higher coreness is strengthened, and vice versa.

The vanilla MSA in the last layer can be redesigned by adopting the coreness of the patches, i.e., injecting the learned corners into the self-attention weights of the class token. As a result, the coreness aware self-attention (CSA) in the last transformer layer is defined as:
\begin{equation}\label{SALastLayer}
CSA(q,K,V) = softmax(\frac{qK^{T}}{\sqrt{d} }  )\odot \left [ 1;C\left (  K_{patch}\right )  \right ] V
\end{equation}
where $q$ is the query of the class token, $K_{patch}$ is the key of the patch tokens, $\odot$ is the dot product, and $\left [ ; \right ] $ is the  concatenation operation. Obviously, the CSA means that the class token takes more information from dominant-invariant patches with high coreness and hinders information from patches with low coreness. The coreness aware multi-head self-attention is therefore defined as:
\begin{equation}\label{C-MSA}
C\text{-}MSA(q,K,V)=Concat(head_{1},...,head_{k} )W^{O}
\end{equation}
where $head_{i}=CSA\left ( qW_{i}^{q}, KW_{i}^{K} , VW_{i}^{V}   \right )  $. Taken them together, the operations in the last transformer layer are formulated as:
\begin{equation}\label{OpeLL}
\begin{split}
& \hat{z}^{l} =C\text{-}MSA\left ( LN\left ( z^{l-1}  \right )  \right ) +z^{l-1}\\
& z^{l}=MLP\left ( LN\left ( \hat{z}^{l} \right )  \right ) + \hat{z} ^{l} 
\end{split}
\end{equation}
In this way, the core-periphery aware transformer layer focuses on fine-grained features that are dominant-invariant and are discriminative for classification. Here $l=L$, $L$ is the number of transformer layers in ViT architecture.

\subsubsection{ Embedding Fusion on Dual-Domain}
More and more evidence shows that adding perturbations enhances model robustness \cite{Sun22} \cite{Pereira21}~\cite{YuNoisyNN2023}. To enhance the stability and robustness of the generated CP graphs and to make the model resistant to noisy perturbations, we incorporate specific perturbations, as outlined in~\cite{YuNoisyNN2023}, into the core-periphery aware transformer layer. Actually, adding single-layer perturbation imposes a regularization on multiple layers simultaneously \cite{Sun22}.

Given an image $x_{i}$ either in the source domain or target domain, let $b_{x_{i}}$ be its input token sequence at the core-periphery aware transformer layer. $b_{x_{i}}$ is viewed as a representation of $x_{i}$ in the latent space. It is not effective to perturb the $x_{i}$ arbitrarily since its dimension in the latent space is high while its support data is limited. Thus we use the next token sequences $b_{x_{j}}, j \in \{1,..., B\}$ in the batch from the same domain to implement embedding fusion~\cite{YuNoisyNN2023}, where $B$ is the batch size. The perturbed token sequence of $b_{x_{i}}$ is represented as:
\begin{equation}\label{perturbation}
\tilde{b}_{x_{i}} = {b}_{x_{i}} + \mu \left [ {b}_{x_{j}} - {b}_{x_{i}} \right ]_{\times } , i\ne j
\end{equation}
where $\mu$ is the perturbation strength, and ${\times }$ means no gradient. The perturbed latent representations aid in generating robust CP graphs and improve performance by effectively reducing task entropy, as elaborated in~\cite{YuNoisyNN2023}. \textbf{Note that the Eq. \ref{perturbation} utilized in this study does not represent the optimal choice for enhancing UDA performance; for the optimal embedding fusion strategy, please refer to \cite{YuNoisyNN2023}.}

\subsection{Core-Periphery Guided Transformer Layer}
With the representation paradigm, an unweighted complete graph can represent the self-attention of the vanilla ViT, and similarly, the core-periphery constraints can be effectively and conveniently infused into the ViT architecture by upgrading the complete graph with the generated weighted CP graphs, which is illustrated in the right part of Figure \ref{model}. Remember the generation process of the CP graph in the previous section, with the guidance of the CP graph, the first $L-1$ transformer layer will focus on the likely core patches, i.e., dominant-invariant features, and suppress the information flow among periphery patches.

A CP graph can be represented by $ \mathcal{G} = ( \mathcal{V}, \mathcal{E})$, with nodes set $ \mathcal{V}= \{ {\nu}_{1},..., {\nu}_{n} \}$, edges set $ \mathcal{E} \subseteq \{ ({\nu}_{i}, {\nu}_{j} )| {\nu}_{i}, {\nu}_{j} \in \mathcal{V} \} $, and adjacency matrix $M_{cp}$. The CP graph guided self-attention for a specific patch $i$ at $r$-th layer of RCCT is defined as:
\begin{equation}\label{patchCP}
x_{i}^{(r+1)}= \sigma^{(r)}( \{ (\frac{q_{i}^{(r)} (K_{j}^{(r)})^{T}  }{\sqrt{d_{k} } } )V_{j}^{(r)}, {\forall} j \in N(i) \} )
\end{equation}
where $\sigma(\cdot)$ is the activation function, which is usually the softmax function in ViTs, $q_{i} ^ {(r)} $ is the query of patches in the $i$-th node in $ \mathcal{G} $, $N(i)=\{i|i \vee (i,j)\in \mathcal{E}\}$ are the neighborhood nodes of node $i$, $d_k$ is the dimension of queries and keys, and $K_{j}^{(r)}$ and $V_{j}^{(r)}$ are the key and value of patches in node $j$. Therefore, the CP graph guided self-attention that is conducted at
the patch level can be formulated as:
 \begin{equation}\label{PatchCPSA}
Attention(Q,K,V,M_{cp})=softmax(\frac{QK^T\odot M_{cp}}{\sqrt{d_k} }V )
\end{equation}
where queries, keys, and values of all patches are packed into matrices $Q$, $K$, and $V$, respectively, $M_{cp}$ is the adjacency matrix provided by the last transformer layer.  Similar to the multi-head attention in transformers, our proposed CP-guided multi-head attention is formulated as:
\begin{equation}
\begin{aligned}
MSA(Q,K,V,M_{cp})=Concat(head_{1}, ...,head_{h})W^{o}\\
\end{aligned}
\end{equation}
where $ head_{i}=Attention(  QW_{i}^{Q}, KW_{i}^{K} , VW_{i}^{V},M_{cp} ) $
where the parameter matrices $W_{i}^{Q}$, $W_{i}^{K}$, $W_{i}^{V}$ and $W^{O}$ are the projections. Multi-head attention helps the model to jointly aggregate information from different representation subspaces at various positions. In this work, we apply the CP constraints to each representation subspace.

\begin{table*}[t]
\caption{ Comparison with SOTA methods on \textbf{Office-Home}. The best performance is marked in red.} 
\centering
\setlength{\tabcolsep}{1.3mm}{} 
\begin{tabular}{ cccccccccccccc }
\hline
Method        & Ar2Cl & Ar2Pr & Ar2Re & Cl2Ar & Cl2Pr & Cl2Re & Pr2Ar & Pr2Cl & Pr2Re & Re2Ar & Re2Cl & Re2Pr & Avg. \\ \hline
ResNet-50\cite{he2016deep}     & 44.9                & 66.3                & 74.3                & 51.8                & 61.9                & 63.6                & 52.4                & 39.1                & 71.2                & 63.8                & 45.9                & 77.2                & 59.4 \\ 
MinEnt\cite{Grandvalet04}        & 51.0                & 71.9                & 77.1                & 61.2                & 69.1                & 70.1                & 59.3                & 48.7                & 77.0                & 70.4                & 53.0                & 81.0                & 65.8 \\ 
SAFN\cite{XuICCV19}          & 52.0                & 71.7                & 76.3                & 64.2                & 69.9                & 71.9                & 63.7                & 51.4                & 77.1                & 70.9                & 57.1                & 81.5                & 67.3 \\ 
CDAN+E\cite{Long18}        & 54.6                & 74.1                & 78.1                & 63.0                & 72.2                & 74.1                & 61.6                & 52.3                & 79.1                & 72.3                & 57.3                & 82.8                & 68.5 \\ 
DCAN\cite{LiAAAI20}          & 54.5                & 75.7                & 81.2                & 67.4                & 74.0                & 76.3                & 67.4                & 52.7                & 80.6                & 74.1                & 59.1                & 83.5                & 70.5 \\ 
BNM \cite{CuiCVPR20}          & 56.7                & 77.5                & 81.0                & 67.3                & 76.3                & 77.1                & 65.3                & 55.1                & 82.0                & 73.6                & 57.0                & 84.3                & 71.1 \\ 
SHOT\cite{Liang20}          & 57.1                & 78.1                & 81.5                & 68.0                & 78.2                & 78.1                & 67.4                & 54.9                & 82.2                & 73.3                & 58.8                & 84.3                & 71.8 \\ 
ATDOC-NA\cite{LiangCVPR21}      & 58.3                & 78.8                & 82.3                & 69.4                & 78.2                & 78.2                & 67.1                & 56.0                & 82.7                & 72.0                & 58.2                & 85.5                & 72.2 \\ \hline
ViT-B\cite{dosovitskiy2020image}         & 54.7                & 83.0                & 87.2                & 77.3                & 83.4                & 85.6                & 74.4                & 50.9                & 87.2                & 79.6                & 54.8                & 88.8                & 75.5 \\
TVT-B\cite{Yang23}         & 74.9                & 86.8                & 89.5                & 82.8                & 88.0                & 88.3                & 79.8                & 71.9                & 90.1                & 85.5                & 74.6                & 90.6                & 83.6 \\ 
CDTrans-B\cite{Xu22}     & 68.8                & 85.0                & 86.9                & 81.5                & 87.1                & 87.3                & 79.6                & 63.3                & 88.2                & 82.0                & 66.0                & 90.6                & 80.5 \\ 
SSRT-B \cite{Sun22}       & 75.2                & 89.0                & 91.1                & 85.1                & 88.3                & 90.0                & 85.0                & 74.2                & 91.3                & 85.7                & 78.6                & 91.8                & 85.4 \\
CCT-B (ours)  & 77.6                & 89.6                & 90.7                & 85.0                & 89.3                & 89.7                & 84.4                & 74.6                & 91.9                & 86.6                & 77.0                & 91.8                & 85.7 \\ 
RCCT-B (ours) &  {\color{red}\textbf{80.1}}   &  {\color{red}\textbf{91.4}}     & {\color{red}\textbf{92.9}}      &  {\color{red}\textbf{87.9}}       &  {\color{red}\textbf{92.2}}                & {\color{red}\textbf{92.2}}     & {\color{red}\textbf{86.3}}    & {\color{red}\textbf{79.5}}             & {\color{red}\textbf{93.1}}   &  {\color{red}\textbf{88.9}}   &  {\color{red}\textbf{81.0}}                &  {\color{red}\textbf{93.8}}    & {\color{red}\textbf{88.3}} \\ \hline
\end{tabular}
\label{Office-Home}
\end{table*}

\begin{table*}[t]
\caption{ Comparison with SOTA methods on \textbf{Visda2017}. The best performance is marked in red.}
\centering
\setlength{\tabcolsep}{1.9mm}{} 
\begin{tabular}{ cccccccccccccc }
\hline
Method     & plane & bcycl & bus & car & horse & knife & mcycl & person & plant & sktbrd & train & truck & Avg. \\ \hline
ResNet-50\cite{he2016deep}     & 55.1                & 53.3              & 61.9              & 59.1               &80.6                &17.9             &  79.7               & 31.2                &  81.0               & 26.5                &  73.5               & 8.5                & 52.4 \\ 
DANN\cite{Ganin15}          & 81.9              & 77.7                &  82.8                & 44.3                &  81.2                & 29.5                &  65.1                & 28.6                &  51.9              & 54.6                & 82.8                &7.8                & 57.4\\ 
MinEnt\cite{Grandvalet04}        & 80.3               & 75.5                & 75.8               &48.3                &  77.9                & 27.3                & 69.7                & 40.2               & 46.5                & 46.6               & 79.3                &16.0                & 57.0 \\ 
SAFN\cite{XuICCV19}          &93.6                & 61.3                & 84.1               & 70.6               & 94.1               & 79.0               &  91.8                & 79.6               &  89.9                & 55.6                &  89.0              & 24.4              & 76.1 \\ 
CDAN+E\cite{Long18}        & 85.2               &  66.9              & 83.0                &  50.8              & 84.2                & 74.9                & 88.1               &74.5                & 83.4             &76.0              &81.9              &  38.0                & 73.9 \\ 
BNM \cite{CuiCVPR20}          &89.6            & 61.5                &  76.9                & 55.0             & 89.3             & 69.1                & 81.3             & 65.5            &  90.0               & 47.3              & 89.1            & 30.1              &70.4 \\ 
CGDM\cite{DuCVPR21}      & 93.7              &82.7               & 73.2             & 68.4               & 92.9                & 94.5              &88.7                & 82.1             &93.4                & 82.5                & 86.8                & 49.2               & 82.3 \\ 
SHOT\cite{Liang20}          &94.3               & 88.5               & 80.1                 & 57.3              &  93.1             &  93.1              & 80.7               & 80.3              & 91.5               &89.1             &  86.3           & 58.2            & 82.9 \\ 
\hline
ViT-B\cite{dosovitskiy2020image}         & 97.7              & 48.1                & 86.6              & 61.6               & 78.1              &  63.4               & 94.7                & 10.3               & 87.7            &   47.7                & 94.4                &  35.5              & 67.1 \\
TVT-B\cite{Yang23}         &   92.9           & 85.6               &77.5               & 60.5              &  93.6             & 98.2            &89.4                & 76.4               & 93.6               & 92.0                & 91.7                & 55.7               & 83.9 \\ 
CDTrans-B\cite{Xu22}     & 97.1                & 90.5               &82.4              & 77.5               & 96.6             & 96.1               &  93.6                &{\color{red}\textbf{88.6}}               &  {\color{red}\textbf{97.9}}                & 86.9               &  90.3             & {\color{red}\textbf{62.8}}             & 88.4 \\ 
SSRT-B \cite{Sun22}       & {\color{red}\textbf{98.9}}               & 87.6              & {\color{red} \textbf{89.1} }         &{\color{red} \textbf{84.8} }     & 98.3              & {\color{red}\textbf{98.7} }              &{\color{red}\textbf{96.3} }               & 81.1               &94.9             & {\color{red}\textbf{97.9} }          & 94.5                & 43.1               & 88.8 \\
CCT-B (ours)  & 97.1                & 92.9            & 78.0            & 64.1                & 97.5                &96.5 & 90.6            & 78.0             & 91.2              & 95.6            & 93.8                & 65.6             & 86.7 \\ 
RCCT-B (ours) &  98.4   &  {\color{red}\textbf{95.9}}     &87.7    &  77.3     &  {\color{red}\textbf{98.9}}                & 96.7     & 95.8    & 82.6    & 96.4   &  {\color{red}\textbf{97.9}}   &  {\color{red}\textbf{97.8}}                &  {\color{red}\textbf{62.8}}    & {\color{red}\textbf{90.7}} \\ \hline
\end{tabular}
\label{Visda17}
\end{table*}

\subsection{Overall Objective Function }
Since our proposed RCCT has a classifier, a self-clustering module, a patch discriminator, and a global discriminator, there are four terms in the overall objective function. The classification loss term is formulated as:
\begin{equation}\label{clcloss}
L_{clc} \left ( x^{s}, y^{s}   \right ) = \frac{1}{n_{s}}\sum_{x_{i}\in D_{s}  }  L_{CE} \left ( G_{c} \left ( G_{f} \left ( x_{i}^{s} \right )  \right ) , y_{i}^{s} \right ) 
\end{equation}
where $G_{c}$ is the classifier.

The domain discriminator takes the class token and tries to discriminate the class token, i.e., the representation of the entire image, to the source or target domain. The domain adversarial loss term is formulated as:
\begin{equation}\label{domainloss}
L_{dis}\left ( x^{s}, x^{t}  \right )  = -\frac{1}{n} \sum_{x_{i} \in D}L_{ce}\left ( D_{g}\left ( G_{f}\left ( x_{i}^{*} \right ) , y_{i}^{d} \right )  \right )  
\end{equation}
where $D{g}$ is the domain discriminator, and $y_{i}^{d}$ is the the domain label ((i.e., $y_{i}^{d}= 1$ means source domain, $y_{i}^{d}= 0$ is target).

The self-clustering module is inspired by the cluster assumption \cite{Chapelle05} and the probability $p^{t}=softmax\left ( G_{c}\left ( G_{f}\left ( x^{t} \right )  \right )  \right ) $ of target image $x_{t}$ is optimized to maximize the mutual information with $x_{t}$ \cite{Yang23}. The self-clustering loss term is formulated as:
\begin{equation}\label{self-cluseringloss}
I\left ( p^{t}; x^{t} \right ) = H\left ( \bar{p^{t}}  \right ) -\frac{1}{n_{t}} \sum_{i=1}^{n_{t}}H\left ( p_{i}^{t} \right )  
\end{equation}
where $p_{i}^{t}=softmax\left ( G_{c}\left ( G_{f}\left ( x_{i}^{t} \right )  \right )  \right )$ and $\bar{p^{t}}=\mathbb{E}\left [ p^{t} \right ] $. The self-clustering loss encourages the model to learn clustered target features 

Take classification loss (Eq. \ref{clcloss}), domain adversarial loss (Eq. \ref{domainloss}), patch adversarial loss (Eq. \ref{PatchLevelloss}), and self-clustering loss (Eq. \ref{self-cluseringloss}) together, the overall objective function is therefore formulated as:
\begin{equation}
L_{clc} \left ( x^{s}, y^{s}   \right ) + \alpha L_{dis}\left ( x^{s}, x^{t}  \right ) + \beta L_{pat}\left ( x^{s},x^{t} \right ) - \gamma I\left ( p^{t}; x^{t} \right )
\end{equation}
where $\alpha$, $\beta$, and $\gamma$ are the hyperparameters that control the influence of subterms on the overall function. 

\begin{table*}[!htp] \footnotesize
\caption{ Comparison with SOTA methods on \textbf{DomainNet}. The best performance is marked in red.} 
\begin{subfloatrow}
\setlength{\tabcolsep}{0.8mm}{} 
\begin{tabular}[t]{cccccccc}
  \hline
 \makecell[c]{ResNet\\101 \cite{he2016deep}} & clp &  inf & pnt & qdr & rel &  skt & Avg.
  \\
  \hline
 clp & - & 19.3 & 37.5 & 11.1 & 52.2 & 41.1 & 32.2   \\
  \hline
 inf & 30.2 & - & 31.2 & 3.6 & 44.0 & 27.9 & 27.4   \\
  \hline
 pnt & 39.6 & 18.7 & - & 4.9 & 54.5 & 36.3 & 30.8  \\
  \hline
 qdr & 7.0 & 0.9 & 1.4 & - & 4.1 & 8.3 & 4.3  \\
  \hline
 rel & 48.4 & 22.2 & 49.4 & 6.4 & - & 38.8 & 33.0  \\
   \hline
 skt & 46.9 & 15.4 & 37.0 & 10.9 & 47.0 & - &  31.4  \\
  \hline
Avg. & 34.4 & 15.3 & 31.3 & 7.4 & 40.4 & 30.5 & 26.6  \\
  \hline
\end{tabular}

\begin{tabular}[t]{cccccccc}
  \hline
 \makecell[c]{MIMTFL\\\cite{GaoECCV20}} & clp &  inf & pnt & qdr & rel &  skt & Avg.
  \\
  \hline
  clp & - & 15.1 & 35.6 & 10.7 & 51.5& 43.1 & 31.2   \\
  \hline
 inf & 32.1 & - & 31.0 & 2.9 & 48.5 & 31.0 & 29.1   \\
  \hline
 pnt & 40.1 & 14.7 & - & 4.2 & 55.4 & 36.8 & 30.2  \\
  \hline
 qdr & 18.8 & 3.1 & 5.0 & - & 16.0 & 13.8 & 11.3  \\
  \hline
 rel & 48.5 & 19.0 & 47.6 & 5.8 & - & 39.4 & 22.1  \\
   \hline
 skt & 51.7 & 16.5 & 40.3 & 12.3 & 53.5 & - &  34.9  \\
  \hline
Avg. & 38.2 & 13.7 & 31.9 & 7.2 & 45.0 & 32.8 & 28.1  \\
  \hline
\end{tabular}

\begin{tabular}[t]{cccccccc}
  \hline
  \makecell[c]{CGDM\\ \cite{DuCVPR21} }& clp &  inf & pnt & qdr & rel &  skt & Avg.
  \\
  \hline
  clp & - & 16.9 & 35.3 & 10.8 & 53.5 & 36.9 & 30.7   \\
  \hline
 inf & 27.8 & - & 28.2 & 4.4 & 48.2 & 22.5 & 26.2   \\
  \hline
 pnt & 37.7 & 14.5 & - & 4.6 & 59.4 & 33.5 & 30.0  \\
  \hline
 qdr & 14.9 & 1.5 & 6.2 & - & 10.9 & 10.2 &8.7  \\
  \hline
 rel & 49.4 & 20.8 & 47.2 & 4.8 & - & 38.2 & 32.0  \\
   \hline
 skt & 50.1 & 16.5 & 43.7 & 11.1 & 55.6 & - &  35.4  \\
  \hline
Avg. & 36.0 & 14.0 & 32.1 & 7.1 & 45.5 & 28.3 & 27.2  \\
  \hline
\end{tabular}
\end{subfloatrow}

\begin{subfloatrow}
\setlength{\tabcolsep}{0.6mm}{} 
\begin{tabular}[t]{cccccccc}
  \hline
 \makecell[c]{MDD+SCDA\\ \cite{LiICCV21} } & clp &  inf & pnt & qdr & rel &  skt & Avg.
  \\
  \hline
  clp & - & 20.4 & 43.3 & 15.2 & 59.3 & 46.5 & 36.9   \\
  \hline
 inf & 32.7 & - & 34.5 & 6.3 & 47.6 & 29.2 & 30.1  \\
  \hline
 pnt & 46.4 & 19.9 & - & 8.1 & 58.8 & 42.9 & 35.2  \\
  \hline
 qdr & 31.1 & 6.6 & 18.0 & = & 28.8 & 22.0 & 21.3  \\
  \hline
 rel & 55.5 & 23.7 & 52.9 & 9.5 & - & 45.2 & 37.4  \\
   \hline
 skt & 55.8 & 20.1 & 46.5 & 15.0 & 56.7 & - &  38.8  \\
  \hline
Avg. & 44.3 & 18.1 & 39.0 & 10.8 & 50.2 & 37.2 & 33.3  \\
  \hline
\end{tabular}

\begin{tabular}[t]{cccccccc}
  \hline
 \makecell[c]{ViT-Base \\ \cite{dosovitskiy2020image} } & clp &  inf & pnt & qdr & rel &  skt & Avg.
  \\
  \hline
  clp & - & 27.2 & 53.1 & 13.2 & 71.2 & 53.3 & 43.6   \\
  \hline
 inf & 51.4 & - & 49.3 & 4.0 & 66.3 & 41.1 & 42.4 \\
  \hline
 pnt & 53.1 & 25.6 & - & 4.8 & 70.0 & 41.8 & 39.1  \\
  \hline
 qdr & 30.5 & 4.5 & 16.0 & - & 27.0 & 19.3 & 19.5  \\
  \hline
 rel & 58.4 & 29.0 & 60.0 & 6.0 & - & 45.8 & 39.9  \\
   \hline
 skt & 63.9 & 23.8 & 52.3 & 14.4 & 67.4 & - &  44.4  \\
  \hline
Avg. & 51.5 & 22.0 & 46.1 & 8.5 & 60.4 & 40.3 & 38.1  \\
  \hline
\end{tabular}

\begin{tabular}[t]{cccccccc}
  \hline
 \makecell[c]{CDTrans-B\\ \cite{Xu22} } & clp &  inf & pnt & qdr & rel &  skt & Avg.
  \\
  \hline
  clp & - & 29.4 & 57.2 & 26.0 & 72.6 & 58.1 & 48.7   \\
  \hline
 inf & 57.0 & - & 54.4 & 12.8 & 69.5 & 48.4 & 48.4   \\
  \hline
 pnt & 62.9 & 27.4 & - & 15.8 & 72.1 & 53.9 & 46.4  \\
  \hline
 qdr & 44.6 & 8.9 & 29.0 & - & 42.6 & 28.5 & 30.7  \\
  \hline
 rel & 66.2 & 31.0 & 61.5 & 16.2 & - & 52.9 & 45.6  \\
   \hline
 skt & 69.0 & 29.6 & 59.0 & 27.2 & 72.5 & - &  51.5  \\
  \hline
Avg. & 59.9 & 25.3 & 52.2 & 19.6 & 65.9 & 48.4 & 45.2  \\
  \hline
\end{tabular}
\end{subfloatrow}

\begin{subfloatrow}
\setlength{\tabcolsep}{0.85mm}{} 
\begin{tabular}[t]{cccccccc}
  \hline
  \makecell[c]{SSRT-B\\ \cite{Xu22} } & clp &  inf & pnt & qdr & rel &  skt & Avg.
  \\
  \hline
  clp & - & 33.8 & 60.2 & 19.4 & 75.8 & 59.8 & 49.8   \\
  \hline
 inf & 55.5 & - & 54.0 & 9.0 & 68.2 & 44.7 & 46.3   \\
  \hline
 pnt & 61.7 & 28.5 & - & 8.4 & 71.4 & 55.2 & 45.0  \\
  \hline
 qdr & 42.5 & 8.8 & 24.2 & - & 37.6 & 33.6 & 29.3  \\
  \hline
 rel & 69.9 & 37.1 & 66.0 & 10.1 & - & 58.9 & 48.4  \\
   \hline
 skt & 70.6 & 32.8 & 62.2 & 21.7 & 73.2 & - &  52.1  \\
  \hline
Avg. & 60.0 & 28.2 & 53.3 & 13.7 & 65.3 & 50.4 & 45.2  \\
  \hline
\end{tabular}

\begin{tabular}[t]{cccccccc}
  \hline
 \makecell[c]{CCT-B \\(ours) } & clp &  inf & pnt & qdr & rel &  skt & Avg.
  \\
  \hline
  clp & - & 30.6 & 56.9 & 17.8 & 69.8 & 58.0 & 46.6   \\
  \hline
 inf & 53.9 & - & 47.6 & 9.3 & 69.2 & 45.0 & 45.0   \\
  \hline
 pnt & 52.5 & 26.2 & - & 8.4 & 70.0 & 48.0 & 41.0  \\
  \hline
 qdr & 37.6 & 10.5 & 19.6 & - & 29.3 & 26.9& 24.8  \\
  \hline
 rel & 63.9 & 32.4 & 61.7 & 11.6 & - & 53.4 & 44.6  \\
   \hline
 skt & 67.3 & 28.9 & 60.0 & 20.5 & 71.5 & - &  49.6  \\
  \hline
Avg. & 55.0 & 25.7 & 49.2 & 13.5 & 62.0 & 46.3 & 42.0  \\
  \hline
\end{tabular}

\begin{tabular}[t]{cccccccc}
  \hline
 \makecell[c]{RCCT-B \\(ours) } & clp &  inf & pnt & qdr & rel &  skt & Avg.
  \\
  \hline
  clp & - & 32.4 & 60.2 & 21.1 & 78.5 & 63.2 & 51.1   \\
  \hline
 inf & 57.5 & - & 55.8 & 9.7 & 71.6 & 47.8 & 48.5   \\
  \hline
 pnt & 63.5 & 29.4 & - & 9.4 & 72.5 & 54.9 & 45.9  \\
  \hline
 qdr & 42.2 & 12.4 & 23.6 & - & 33.8 & 30.6 & 28.5  \\
  \hline
 rel & 70.4 & 34.3 & 67.3 & 12.9 & - & 57.8 & 48.5  \\
   \hline
 skt & 72.6 & 31.9 & 64.1 & 22.1 & 75.4 & - &  53.2  \\
  \hline
Avg. & 61.2 & 28.1 & 54.2 & 15.0 & 66.4 & 50.9 & {\color{red}\textbf{46.0}}  \\
  \hline
\end{tabular}
\end{subfloatrow}
\label{DomainNet}
\end{table*}

\begin{table}[t] \small
\caption{ Comparison with SOTA methods on \textbf{Office-31}. The best performance is marked in red.} 
\centering
\setlength{\tabcolsep}{0.9mm}{} 
\begin{tabular}{cccccccc}
\hline
Method     & A2W  & D2W  & W2D   & A2D  & D2A  & W2A  & Avg. \\ 
ResNet-50\cite{he2016deep}   & 68.4 & 96.7 & 99.3  & 68.9 & 62.5 & 60.7 & 76.1 \\ 
DANN\cite{Ganin15}        & 82.0 & 96.9 & 99.1  & 79.7 & 68.2 & 67.4 & 82.2 \\ 
rRGrad+CAT\cite{DengICCV19} & 94.4 & 98.0 & 100.0 & 90.8 & 72.2 & 70.2 & 87.6 \\ 
SAFN+ENT\cite{XuICCV19}   & 90.1 & 98.6 & 99.8  & 90.7 & 73.0 & 70.2 & 87.1 \\ 
CDAN+TN \cite{WangNIPS19}   & 95.7 & 98.7 & 100.0 & 94.0 & 73.4 & 74.2 & 89.3 \\ 
TAT \cite{LiuICML19}       & 92.5 & 99.3 & 100.0 & 93.2 & 73.1 & 72.1 & 88.4 \\ 
SHOT \cite{Liang20}      & 90.1 & 98.4 & 99.9  & 94.0 & 74.7 & 74.3 & 88.6 \\ 
MDD+SCDA\cite{LiICCV21}   & 95.3 & 99.0 & 100.0 & 95.4 & 77.2 & 75.9 & 90.5 \\ \hline
ViT-B\cite{dosovitskiy2020image}      & 91.2 & 99.2 & 100.0 & 93.6 & 80.7 & 80.7 & 91.1 \\ 
TVT-B \cite{Yang23}     & 96.4 & 99.4 & 100.0 & 96.4 & 84.9 & 86.1 & 93.9 \\ 
CDTrans-B \cite{Xu22}  & 96.7 & 99.0 & 100.0 & 97.0 & 81.1 & 81.9 & 92.6 \\ 
SSRT-B \cite{Sun22}    & {\color{red}\textbf{97.7}} & 99.2 & 100.0 & {\color{red}\textbf{98.6}} & 83.5 & 82.2 & 93.5 \\ 
CCT(ours)        & 96.0 & {\color{red}\textbf{99.5}} & 100.0 & 94.4 & 84.5 & 85.1 & 93.3 \\ 
RCCT(ours)       & 97.4 & {\color{red}\textbf{99.5}} & 100.0 & 96.4 & {\color{red}\textbf{88.1}} & {\color{red}\textbf{88.7}} & {\color{red}\textbf{95.0}} \\ \hline
\end{tabular}
\label{office31}
\end{table}

\section{Experiments}
We evaluate our proposed RCCT on widely used UDA benchmarks, including \textbf{Office-31} \cite{Saenko10}, \textbf{Office Home} \cite{Venkateswara17}, \textbf{VisDA2017} \cite{Peng17}, and \textbf{DomainNet} \cite{Peng19}. \textbf{Office-31} has 3 domains, i.e., Amazon (A), DSLR (D), and Webcam (W). \textbf{Office-Home} has 4 domains: Artistic (Ar), Clip Art (Cl), Product (Pr), and Real-world (Rw) images. \textbf{VisDA2017} has 12 classes. \textbf{DomainNet} has 6 domains: Quickdraw (qdr), Real (rel), Sketch (skt), Clipart (clp), Infograph (inf), Painting (pnt). A detailed description of datasets is provided in the supplementary.

We use the ViT-base with a 16×16 patch size  (ViT-B/16) \cite{dosovitskiy2020image} \cite{Steiner21}, pre-trained on ImageNet \cite{Russakovsky15}, as our vision transformer backbone. The ViT-B/16 contains 12 transformer layers in total. We use minibatch Stochastic Gradient Descent (SGD) optimizer \cite{Ruder18} with a momentum of 0.9 as the optimizer. The batch size is set to 32 for all the experiments. We initialized the learning rate as 0 and linearly warm up to 0.06 after 500 training steps. We then schedule it using the cosine decay strategy. For small to middle-scale datasets Office-31 and Office-Home, the epoch is set to 5000. For large-scale datasets Visda-2017 and DomainNet, the epoch is set to 20000. The perturbation ratio $\mu$ is randomly chosen from $[0,0.5]$. The hyper-parameters $\alpha$, $\beta$, and $\gamma$ are set to $[1.0, 0.01, 0.1]$ for Office-31 and Office-Home, $[0.1, 0.1, 0.1]$ for Visda-2017 and DomainNet.

\begin{figure*}[htp]
\begin{center}
\includegraphics[scale=0.24]{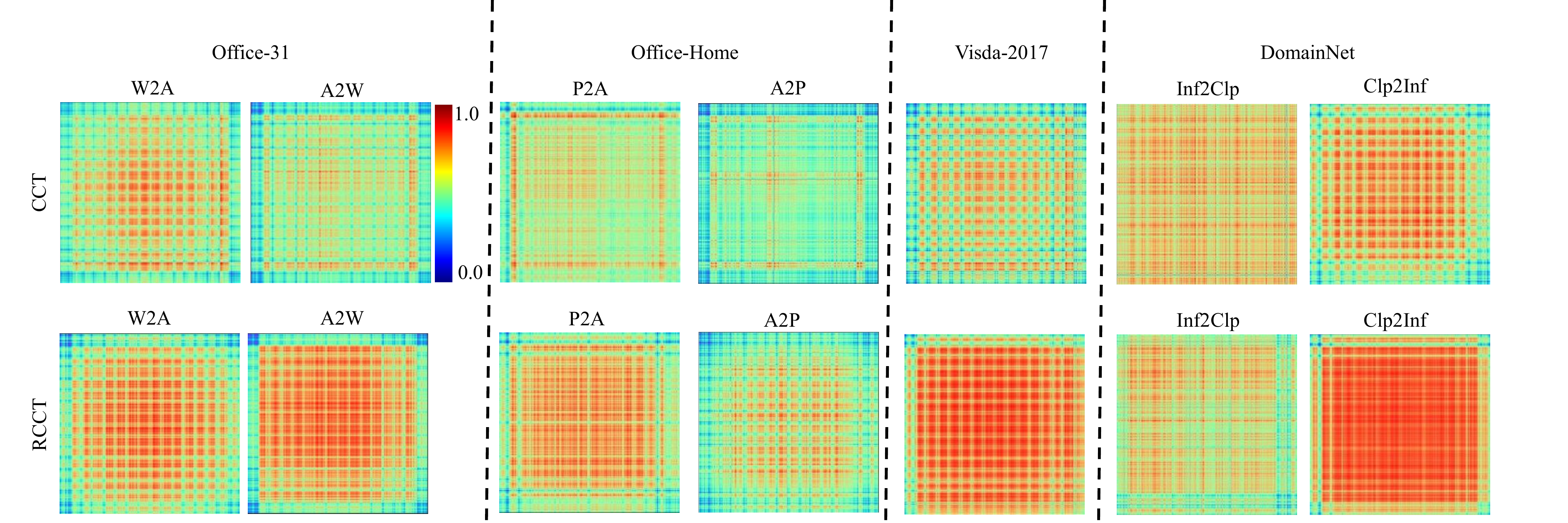}
\end{center}
\caption{The learned core-periphery graphs (adjacency matrics) from some randomly selected domain adaptation tasks. The first line includes the CP graphs generated from the CCT model, while the second lines are the CP graphs from RCCT mode. The texts above the CP graphs show the task to which the CP graphs belong. The redder colors the higher weight.} 
\label{cpgraphs}
\end{figure*}

\subsection{Results}
Table \ref{Office-Home} presents evaluation results on the dataset Office-Home. The “-B” indicates results using ViT-base backbones. RCCT means robust core-periphery constrained transformer, whereas CCT means omitting LFI operation. The methods above the black line are based on CNN architecture, while those under the black line are developed from the Transformer architecture. Up till the present moment, most of the methods for UDA use CNN-based models. It is worth mentioning that our proposed RCCT significantly outperforms all the methods in each domain adaptation task as well as the average results. Table \ref{Visda17} shows results on the dataset VisDA2017. We can observe that the RCCT achieves better performance on average results and in most adaptation tasks. The experimental results on the large dataset DomainNet are shown in Table \ref{DomainNet}. The core-periphery principle enables the model to outperform the ViT baseline, while the LFI operation enhances the model's performance compared to the most advanced methods. The RCCT also achieved the SOTA results on Office-31, as shown in Table \ref{office31}. From the comparisons, the transformer-based methods gain much better results than CNN-based models thanks to their strong transferable feature representations. Compared with other methods, our RCCT-B performs the best on Office-Home, Office-31, DomainNet, and VisDA2017. Even the weaker version of CCT-B can surpass most methods on these datasets.
RCCT-B improves 13.8\% on Office-Home, 3.9\% on Office-31, 23.6\% on VisDA-2017, and 7.9\% on DomainNet, over ViT-B despite that ViT-B baseline is already very competitive. 

\subsection{Learned Core-Periphery Graphs}
Our method adaptively learns core-periphery graphs for different datasets and tasks and uses the learned core-periphery graphs to reschedule the self-attention as defined in Eq. \ref{CPgraphGeneration} \ref{SALastLayer} \ref{PatchCPSA}. Some randomly selected learned core-periphery graphs are shown in Figure \ref{cpgraphs}. The first line includes the learned CP graphs from CCT, while the second line shows the CP graphs learned from RCCT. These graphs have been examined by core-periphery detection algorithms\cite{Kojaku17}. Obviously, the CP graphs from RCCT show more dense and weighted patterns, which helps capture the core patches from different domains and improves domain adaptivity across domains.

\begin{figure}[htp]
\begin{center}
\includegraphics[scale=0.19]{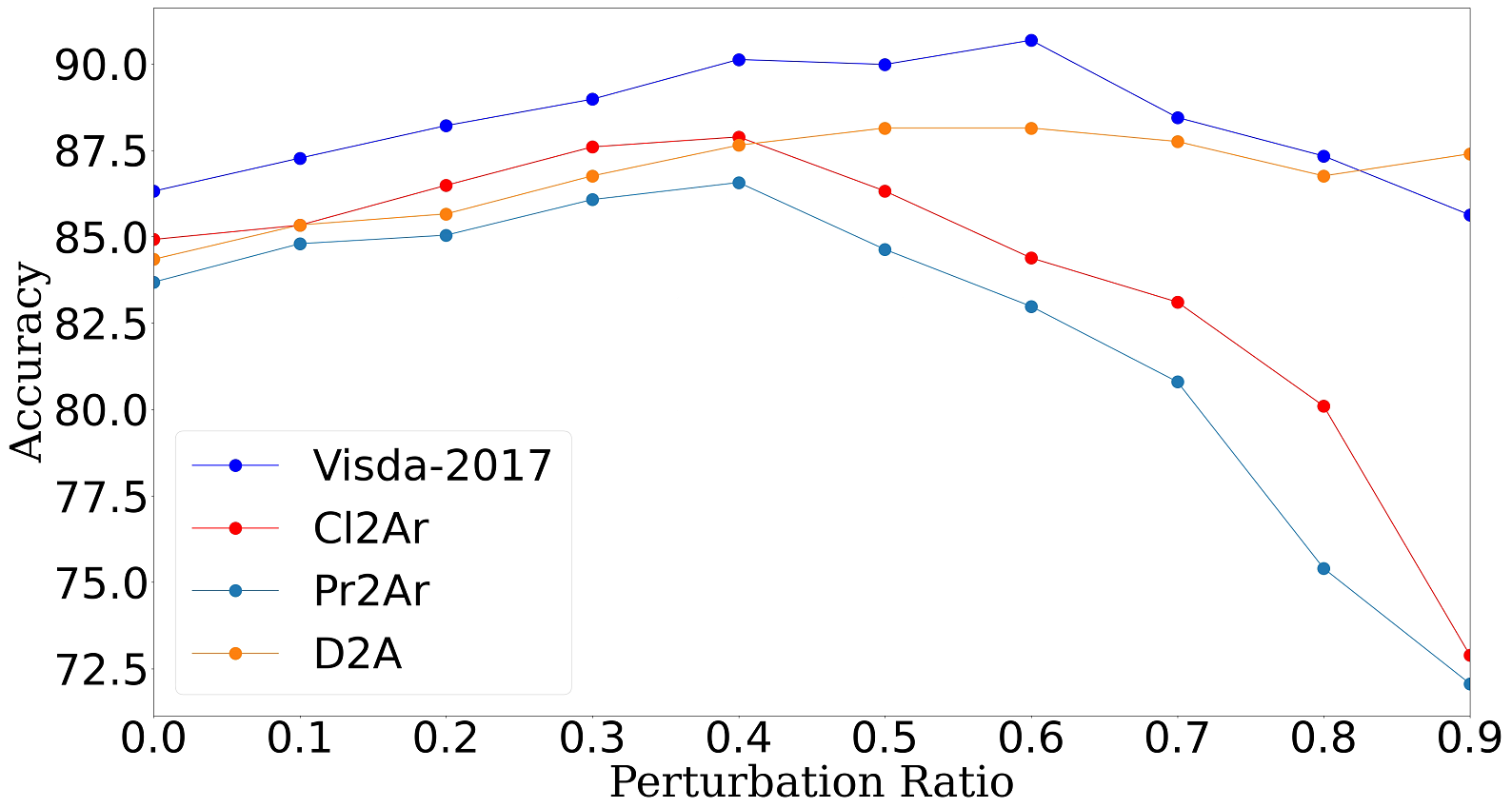}
\end{center}
\caption{The influence of perturbation ratio $\mu$ on accuracy. The perturbation ratio ranges from [0.0, 0.9]. } 
\label{PR}
\end{figure}

\subsection{Ablation Studies}
For all four datasets, we presented the experimental results with and without perturbation, as shown in the last two terms in Table \ref{Office-Home}, \ref{Visda17}, \ref{DomainNet}, \ref{office31}. We also conduct ablation studies on the perturbation ratio. Figure \ref{PR} plots the influence of the perturbation ratio on accuracy. Note when the perturbation ratio is 0, which means there is no perturbation, the RCCT is degraded to CCT. From Figure \ref{PR}, we can observe that RCCT can gain prediction improvements on a wide range of perturbation ratios. For visda2017 and D2A of office-31, the RCCT performs well even if the perturbation ratio is higher than 0.5. We further evaluate the influence of the CP constraints on the model. The results are shown in the supplementary. 

\section{Conclusion}
In this paper, we propose a novel brain-inspired approach, named RCCT for unsupervised domain adaptation. It practically instills the core-periphery constraints into the self-attention in the Transformer architecture. The RCCT can adaptively learn core-periphery graphs by measuring the coreness of patches via a patch discriminator. At the same time, deliberate perturbations are added to force the model to learn robust CP graphs. We use the learned CP graphs to manipulate self-attention weights to strengthen the information communication among higher coreness patches while suppressing that among low coreness patches. Our RCCT achieves competitive results on four popular UDA datasets, outperforming previous methods on some datasets, such as Visda2017. Our work demonstrates that brian-inspired DNNs have promising applications in the UDA.

{\small
\bibliographystyle{ieee_fullname}
\bibliography{egbib}
}

\end{document}